# Exceeding the solar cell Shockley-Queisser limit via thermal up-conversion of low-energy photons

Svetlana V. Boriskina* and Gang Chen*

*Department of Mechanical Engineering, Massachusetts Institute of Technology, Cambridge, MA, 02139, USA*
*e-mails:* [sborisk@mit.edu](sborisk@mit.edu) , [gchen2@mit.edu](gchen2@mit.edu)

**Abstract**

Maximum efficiency of ideal single-junction photovoltaic (PV) cells is limited to 33% (for one sun illumination) by intrinsic losses such as band edge thermalization, radiative recombination, and inability to absorb below-bandgap photons. This intrinsic thermodynamic limit, named after Shockley and Queisser (S-Q), can be exceeded by utilizing low-energy photons either via their electronic up-conversion or via thermophotovoltaic (TPV) conversion process. However, electronic up-conversion systems have extremely low efficiencies, and practical temperature considerations limit the operation of TPV converters to the narrow-gap PV cells. Here we develop a *conceptual* design of a hybrid TPV platform, which exploits *thermal up-conversion* of low-energy photons and is *compatible with conventional silicon PV cells* by using spectral and directional selectivity of the up-converter. The hybrid platform offers sunlight-to-electricity conversion efficiency exceeding that imposed by the S-Q limit on the corresponding PV cells across a broad range of bandgap energies, under low optical concentration (1-300 suns), operating temperatures in the range 900-1700K, and in simple flat panel designs. We demonstrate maximum conversion efficiency of 73% under illumination by non-concentrated sunlight. A detailed analysis of non-ideal hybrid platforms that allows for up to 15% of absorption/re-emission losses yields limiting efficiency value of 45% for Si PV cells.

1. **Introduction**

The quest for efficient low-cost solutions for solar energy conversion faces many obstacles, both, fundamental and technical. The major fundamental – or so-called intrinsic – limitations to the PV conversion efficiency stem from the laws of thermodynamics and quantum mechanics [1-6]. They include: (i) the losses due to thermalization of charge carriers generated by absorption of the photons with the energies above the bandgap of the PV material, (ii) the losses caused by the PV cell inability to use the photons with the energies below the bandgap, and (iii) the losses caused by recombination of the light-generated charge carriers. Technical – or extrinsic – limits, such as e.g., low absorption efficiency of the PV material, can be overcome by the proper design of the PV cell and the solar concentrator device. The explored approaches to improve extrinsic limits include using anti-reflecting coatings and back mirrors, texturing the cell surface with random, periodic or nature-inspired non-periodic nano-patterns, coupling incoming radiation into propagating or localized modes within the cell, etc [7-17]. However, even 'ideal' PV cells have maximum intrinsic efficiency of 33% [1, 2, 6, 18-20] for the illumination by non-concentrated sunlight with the AM1.5D [21] energy spectrum.

Several approaches to designing solar converters with efficiencies exceeding the S-Q limit of single-junction PV cells have been proposed. These include engineering multiple-junction [19, 20] and intermediate-band [22, 23] solar cells, which – in the ideal infinite-junction case – can eliminate the thermalization losses, and using concentrated sunlight [19, 24], which helps to counteract the recombinative radiation losses. However, even for the maximum concentration of solar radiation, the S-Q efficiency of a single-junction PV cell does not exceed 44% (for AM1.5D spectrum) [20]. The highest resulting limiting efficiency of the solar energy-to-electricity conversion can theoretically reach 85% (for AM1.5D spectrum), for a PV cell with an infinite number of p-n junctions [1, 20]. Mechanisms of *down-conversion* of high-energy photons into two or more lower-energy photons of [25-28] have been shown to increase the intrinsic thermodynamic efficiency limit by reducing the thermalization losses. An alternative approach to designing high-efficiency solar energy converters is based on coupling thermal and photovoltaic converters [1, 9, 28-39] and has a theoretical efficiency limit of 85%. However, maximum efficiency is only achievable under unrealistic conditions such as temperature of about 2500K, an ideal narrow-band transmission filter, an infinite ratio of the emitting-to-absorbing surface areas of the intermediate absorber, and maximum optical concentration [1, 30]. Thus, all the proposed specific designs of TPV converters rely on large emitter-to-absorber area ratios (larger than 10:1), high levels of optical concentration (over 1000 suns), and high operating temperatures over 2000K [13, 30, 33, 37, 38]. Although Si PV cells would be excellent candidates for the use as a part of a TPV module owing to their commercial availability, mature fabrication technology, and the high S-Q efficiency, they would only be feasible at impractically high temperatures well above 2000 K. As a result, TPV systems are typically limited to using PV cells with narrow bandgaps in the range 0.4-0.75eV [13, 29] such as GaSb [33], PbS [36], HgCdTe [9], and Ge [40] and effectively operate in the regime of *thermal down-conversion* [9, 13, 28]. In practice, it is difficult to achieve high efficiency in low bandgap PV cells due to increased Auger recombination. Hence, the best demonstrated TPV efficiency for solar energy conversion is 1.7% for GaSb PV cells under 4600 suns illumination [33, 41].

*Electronic up-conversion* of the low-energy photons has been explored as an alternative approach to improve the efficiency limit of the PV cells with larger bandgaps such as Si [42-44]. The most common realization of the up-converter-enhanced PV platform exploits sequential excitation of electrons into an excited state via a lower-lying metastabe state. The efficiency of the ideal up-conversion-enhanced PV cell can theoretically reach 63% [43], but commonly used up-converting systems (such as e.g. rare-earth ions) suffer from very low absorption efficiencies, narrow-band absorption spectra, and high non-radiative recombination rates [45-48]. Although recent work on broadband up-conversion holds some promise for the efficiency improvement [49], efficiencies of PV cells with up-conversion experimentally measured to date have been extremely low (in single digits even under monochromatic light illumination [45, 50]).

Here we present a directionally- and spectrally-selective TPV platform with a total solar-to-electricity conversion efficiency exceeding the Shockley-Queisser limit of the PV cell used as a part of the hybrid system, which is based on the mechanism of *thermal up-conversion* of the below-bandgap-energy photons. The proposed hybrid converter has a simple flat-panel geometry with 1:1 ratio of surface areas of the intermediate absorber, emitter and the PV cell and can operate in the temperature range of 900-1700K under low optical concentration levels (1-300 suns). The converter can incorporate PV cells made of conventional wider-bandgap PV materials such as Si and GaAs. We estimate the *intrinsic thermodynamic efficiency limits* of the proposed conceptual energy conversion scheme (i.e., analogs of the PV cell Shockley-Queisser limit) under ideal and realistic conditions. We discuss some approaches to possible practical realization of the up-conversion concept, however, *estimation of the extrinsic efficiency limits is beyond the scope of this manuscript* as such limits will be highly realization-specific.

2. **Thermal up-conversion concept**

We propose a hybrid energy conversion platform based on *thermal up-conversion* of low-energy photons. It consists of at least one bifacial single-junction PV cell and a solar-thermal up-converter – a slab of absorbing material with spectrally- and angularly-selective surfaces at both, sun-illuminated and shadow sides, as shown in Fig. 1. The photons with the energies below the bandgap of the PV cell material are absorbed by the up-converter, and subsequently re-emitted with higher energies towards a PV cell. The sun-illuminated front surface of the up-converter ($SS_F$ in Fig. 1) is highly absorptive to solar radiation but has angularly- and spectrally-restricted emittance. Ideally, the front surface perfectly absorbs and emits photons within the angular range $\theta_s$ of the incoming below-gap solar radiation ($\theta_s \leq \theta_u^F$, $\varepsilon(\theta < \theta_u^F) = \alpha(\theta < \theta_u^F) = 1$; $\theta$ being the polar angle measured from the normal to the surface), and perfectly blocks emittance at larger angles ($\varepsilon(\theta > \theta_u^F) = \alpha(\theta > \theta_u^F) = 0$). For direct solar radiation at the Earth's surface, the minimum acceptance half-angle is $\theta_s^m = 0.267^o$, which can be increased by using optical concentrating systems up to $\theta_s = 90^o$ for maximum concentration of $C = \sin^{-2} \theta_s^m = 46050$ [24].

Spectral and angular selectivity of surface emittance can be used to increase the efficiency limits of PV and TPV systems operated with low optical concentration of sunlight [4, 9, 13, 34, 51, 52]. Previous work has shown that both, angular and spectral selectivity of the surface emittance can be achieved by surface nano-patterning [9, 28, 51, 53] and/or by using multilayer Bragg reflectors [13]. We have also recently developed an alternative way to achieve angular selectivity by enclosing the up-converter into the reflective cavity with a limited angular aperture [54]. The back surface of the up-converter that faces the PV cell ($SS_B$ in Fig. 1) is angularly-isotropic yet spectrally-selective, and is characterized by a high(low) emittance for photons with the energies above(below) the bandgap of the PV cell. The surface absorptance/emittance characteristics of the up-converter and the PV cell are shown in Fig. 2a and are overlapped with the direct terrestrial solar spectrum AM1.5D (total integrated irradiance 900.1 W·m$^{-2}$) and thermal blackbody spectra shown in Fig. 2b. The illuminated front surface of the up-converter has high absorptance in the frequency range of incoming below-bandgap solar radiation ($E_m \leq h\nu \leq E_g$), while the parasitic thermal radiation in the infra-red (IR) frequency range is suppressed. The back surface of the up-converter can emit and absorb efficiently only the above-bandgap radiation, and this radiation is angularly-isotropic ($\theta_u^B = \pi/2$). The ideal case corresponds to the situation when $\alpha_1^F = \alpha_1^B = 1$ and $\alpha_2^F = \alpha_3^F = \alpha_2^B = 0$. To enable proper comparison to the S-Q limit, the PV cell is assumed to act as a perfect blackbody for the above-bandgap photons and to have zero absorptance/emittance at below-bandgap frequencies.

The incoming energy flux of the below-gap photons ($I_s^{E<E_g}$) heats up the up-converter, which emits the temperature-dependent thermal radiation through its front ($I_u^F(T_u,0)$) and back ($I_u^B(T_u,0)$) surfaces with the spectra determined by their emittance properties shown in Fig. 2a. The photons emitted through the back surface of the PV cell due to radiative recombination of the electron-hole pairs ($I_c^B$) are also absorbed by the up-converter and subsequently re-emitted back to the PV cell. The equilibrium temperature of the up-converter ($T_u$) is determined through the energy balance calculations:

$$I_{abs}^{E_m<E<E_g}(T_s,0) + I_c^B(T_c,eV) = I_u^B(T_u,0) + I_u^F(T_u,0), \qquad (1)$$

where the absorbed/emitted energy fluxes are a function of the temperatures of the sun ($T_s$), the up-converter ($T_u$), and the PV cell ($T_c$), as well as of the angular- and spectral absorptance/emittance characteristics of the PV cell and the up-converter $\alpha(E,\theta)$ [1]:

$$I(T,\mu) = \int_0^\infty N_p(E,T,\mu)E\,dE \int_0^{2\pi} d\varphi \int_0^{\pi/2} \alpha(E,\theta)\sin\theta\cos\theta\,d\theta. \qquad (2)$$

For an angularly-selective surface (with emittance and absorptance being step-wise functions of the zenith angle, $\alpha(\theta > \theta_m) = 0$, and isotropic over the azimuth angle $\varphi$), the absorbed/emitted energy flux takes the following form:

$$I(T,\mu) = \pi \sin^2\theta_m \int_0^\infty \alpha(E) N_p(E,T,\mu) E\,dE, \qquad (3)$$

where $\theta_m = \theta_u^F$ on SS$_F$, $\theta_m = \theta_u^B$ on SS$_B$, and the photon distribution function obeys the Bose-Einstein statistics

$$N_p(E,T,\mu) = \frac{2\pi}{h^3 c^2} \frac{E^2}{\exp((E-\mu)/k_B T) - 1}. \qquad (4)$$

In Eq. (4), $c$ is the speed of light, $h$ is the Planck constant, and $k_B$ is the Boltzmann constant. For the thermal (blackbody) emission, chemical potential of radiated light is zero ($\mu = 0$), while for the luminescent emission from the PV cell $\mu = (E_{FC} - E_{FV}) = e \cdot V$ (Fig. 1).

Once the equilibrium temperature of the up-converter is obtained, the up-conversion efficiency can be calculated as the ratio of the energy carried by the up-converted photons to the energy carried by all below-gap photons incident on the up-converter:

$$\eta_u = I_u^B(T_u,0) \big/ I_s^{E<E_g}(T_s,0). \qquad (5)$$

To estimate the overall intrinsic efficiency limit of the hybrid platform, the up-converter energy balance equation (1) must be solved simultaneously with the detailed balance calculations for the PV cell as outlined in the following section because the up-converter temperature also depends on photons emitted from the PV cell, $I_c^B(T_c, eV)$ in Eq. (1).

## 3. Detailed balance calculations of the thermo-photo-voltaic converter efficiency

This intrinsic thermodynamic efficiency limit of a single-junction PV cell is a function of its electronic bandgap energy, and can be determined via detailed balance calculations [2]. These calculations are based on solving the photon flux balance equations of photon absorption in the PV cell, generation of the electron-hole pairs, and their radiative recombination [1, 2, 19]. We adopt the conventional assumptions: (i) absolute absorptance of the cell $\alpha_c(E > E_g) = 1$, (ii) the number of generated electron-hole pairs being equal to the number of absorbed photons, (iii) infinite mobility of charge carriers – so that the electrons and holes quasi-Fermi levels are flat and separated by the energy gap $\mu = (E_{FC} - E_{FV}) = e \cdot V$ (see Fig. 1), and (iv) radiative recombination being the only recombination mechanism.

The charge current in the PV cell is given by the difference in the generation and recombination rates of the electron-hole pairs:

$$J = e \cdot \left(N_c^{in} - N_c^{out}(eV)\right). \tag{6}$$

The incoming photon flux in Eq. (6) includes both, above-the-bandgap solar photons and the above-the-bandgap photons emitted from the back surface of the up-converter: $N_c^{in} = N_s^{E>E_g}(T_s, 0) + N_u^B(T_u, 0)$. The outgoing photon flux through both surfaces of the PV cell $N_c^{out} = N_c^F(T_c, eV) + N_c^B(T_c, eV)$ is luminescent emission due to the radiative recombination of electron-hole pairs. The photon fluxes in (6) are a function of the temperatures $T_s$, $T_u$, and $T_c$, as well as of the angular- and spectral emittance characteristics defined in Fig. 2a [1]:

$$N(T,\mu) = \pi \sin^2 \theta_m \int_0^\infty \alpha(E) N_p(E,T,\mu) dE .\tag{7}$$

The overall efficiency of the hybrid converter is defined as the ratio of the maximum electrical power delivered to the load to the total power of the sunlight ($I_{sol} = C \cdot I_{AM1.5D}$), which includes both above- and below-gap photons:

$$\eta = \max(J \cdot V)/I_{sol} .\tag{8}$$

The maximum power point $(J_m, V_m)$ can be found from the solution of the following equation:

$$d(J \cdot V)/dV = 0 .\tag{9}$$

Note that as both (1) and (9) contain $T_u$ and $eV$ as parameters, they must be solved simultaneously via an iterative process. The temperature of the PV cell was assumed as $T_u = 300K$ in all the following calculations. To access the accuracy of the commonly-used approximation of the solar energy flux as that of the black body at high temperature, we will compare the hybrid device characteristics under illumination by the terrestrial sunlight with the AM1.5D spectrum and by the radiation from the black body at 6000K [2].

## 4. Results and discussion

By solving Eqs. (1) and (9) numerically, we calculate the limiting efficiency of the proposed hybrid converter as a function of the PV cell bandgap energy ($E_g$), the minimum energy of photons absorbed/emitted through the front surface of the up-converter ($E_m$), the up-converter light acceptance angle $\theta_u^F$, and the optical concentration C. The highest intrinsic efficiency level for the non-concentrated sunlight is reached when $\theta_u^F = \theta_s^m$ and $\theta_c^F = \theta_s^m$. In this case, all the incoming sunlight is absorbed by the PV cell and up-converter, while the thermal and luminescent emission through their front surfaces is limited to the angular range subtended by the sun. Figure 3a shows that the maximum efficiency of the hybrid device can reach 76% if the solar spectrum is approximated as a blackbody spectrum at 6000K (blue lines in Fig. 3) and 73% for the AM1.5D spectrum (red lines in Fig. 3). The efficiency maxima are reached at bandgap energy values higher than those for a single PV cell (dashed lines), and much higher than those for conventional TPV platforms (i.e., 0.4-0.75eV). However, the hybrid device efficiency is significantly improved over that of the PV cell in the whole range of bandgap energy values. As seen from Fig. 3b, reaching higher efficiency requires rising the up-converter temperature over 2000K. Nevertheless, for the PV materials of practical interest, the up-converter temperature stays within practical 900-1600K range. As the thermal radiation from the up-converter increases with the rise of its temperature (3), the up-conversion efficiency drops simultaneously (Fig. 3c). As seen in Fig. 3b, high up-converter temperature can be achieved even for illumination with non-concentrated sunlight. Typically, high levels of optical concentration are required to overcome emission losses from a surface of an absorber kept at elevated temperature (see spectra in Fig. 2c). In this hybrid platform, thermal emission losses are suppressed by the angular selectivity of the up-converter front surface that only enables emission into a narrow cone.

Figure 3d shows that the short circuit current $J_{sc}$ is much higher than that of an individual PV cell, resulting in the higher power output of the device. At the same time, our calculations show that the open circuit voltage $V_{oc}$ of the hybrid device is not noticeably different from that of an individual PV cell. $V_{oc}$ and $J_{sc}$ can be calculated from Eq. (6) assuming $J(eV_{oc}) = 0$ or $J_{sc} = J(eV = 0)$, respectively. It should be noted that the use of the 6000K solar spectrum results in the noticeable overestimation of the short circuit current values over those obtained by using the AM1.5D spectrum (Fig. 3d), while the deviations of the efficiency plots are less significant (Fig. 3a). In the following, only the data obtained by using the AM1.5D spectrum are presented.

We will now estimate the performance of the proposed hybrid device under more realistic conditions. Figure 4 shows the device efficiency when $\theta_u^F = 5^o$, $\theta_c^F = \pi/2$ (i.e., only the up-converter front surface has angular-selective properties, e.g., via its enclosure into a reflective cavity with 5-degrees aperture [54]). It can be seen in Fig. 4a that for *C*=1 (blue solid lines in Fig. 4), the efficiency drops significantly from the ideal case, yet remains higher than the S-Q limit (dashed lines in Fig. 4). Thermal radiation losses through the 5-degree angular cone can be counteracted by using optical concentration, although this also increases the operating temperature of the up-converter (Fig. 4b). Still, for the Si PV cell, 96% up-conversion efficiency and 50% overall device efficiency is achieved for reasonable temperatures of about 1600K. This illustrates the significant difference of the current up-conversion approach compared to the TPV designs based on photon down-conversion, as Si PV cells are well-established technology.

Another parameter that can be tuned to increase the device efficiency is the minimum energy of photons ($E_m$) that can be absorbed or emitted by the up-converter front surface (see Fig. 2). Figure 5 demonstrates that even for *C*=1, the efficiency of the hybrid device can be increased by optimizing the value of the absorptance/emittance cut-off energy $E_m$. The increase of $E_m$ limits the emittance spectral range and thus reduces the parasitic thermal emission from the up-converter. It also blue-shifts both the up-conversion (Fig. 5b) and overall peak efficiency values (Fig. 5a) and increases the temperature of the up-converter (Fig. 5b), which, however, stays below 1700K. As illustrated in Fig. 5c, there is a discrete set of optimum $E_m$ values due to the presence of several $H_2O$ absorption bands in the terrestrial solar spectrum. Further increase of $E_m$ will have a detrimental effect on the device efficiency as the losses due to reduced absorption of the sunlight will start to overcome the useful effect of the reduced thermal emission.

In Figs. 3-5, it was assumed that the up-converter has a 100% absorptance ($\alpha_1^F = \alpha_1^B = 1$) within its absorption windows (i.e., for $E_m < E < E_g$ on the front surface, and $E > E_g$ on its back surface), and is perfectly transparent outside this energy range ($\alpha_2^F = \alpha_3^F = \alpha_2^B = 0$). Real materials, however, have limited absorptance/emittance across the whole energy range, and it's important to estimate the effect of imperfect absorption of up-converter on the hybrid device efficiency. To address this issue, in Fig. 6 we investigate the performance of the hybrid device as a function of the losses due to imperfect absorption/emission characteristics of the up-converter surfaces. The labels in Fig. 6 indicate the ratio of absorptance values (in %) within the up-converter absorption windows (see Fig. 2) and emittance within the photon energy ranges where emission should be blocked (i.e., ideal absorptance/emittance characteristics are labeled as 100/0). Clearly, the increased losses reduce the up-conversion efficiency (Fig. 6b, dashed lines) and thus the overall efficiency of the device (Fig. 6a). These losses can be partially compensated by the optical concentration, and the data presented in Fig. 6 have been obtained for the optical concentration of 300 suns. Note that the maximum possible optical concentration in this case is C=348 [24], and we use *C*=300 to account for transmission/absorption losses of the optical concentrator. Under these conditions, even for the up-converter losses level of 85/15 the hybrid device still out-performs the isolated PV cell (the PV cell S-Q limit values are shown as dashed lines in Fig. 6a). The drop in the efficiency limit results from lower equilibrium temperature of the up-converter due to increased leakage of thermal radiation (Fig. 6b). A close inspection of the thermal emission losses reveals that the strongest efficiency decline results from the below-gap emission through the up-converter back surface, which is characterized by angularly-isotropic emittance. An example of the spectral power of the light emitted through the up-converter back surface for the 90/10 level of emittance is shown in the central inset in Fig. 1, where all the photons with energy below bandgap are lost. The losses through the up-converter front surface, on the other hand, are significantly reduced owing to angular selectivity of this surface. Our calculations show that even for the front surface spectral characteristics at the level of 85/15, a high up-conversion efficiency of 46% for Si (resulting in the overall conversion efficiency of about 45%) is maintained provided the back surface losses are kept at the 95/5 level (Fig. 7). The temperature at which the up-converter reaches equilibrium grows with the bandgap energy, and becomes impractically high for GaAs and larger-bandgap PV cell materials (Fig. 6b). However, for the Si solar cell, the up-converter operating temperature remains below 1700K.

## 5. Entropy generation in the hybrid TPV converter

Clear physical insight into the limits imposed on the conversion efficiency of the proposed hybrid platform can be gained from the entropy balance in the system. The solar energy cannot be converted to electricity with 100% efficiency as it is not entropy-free [1, 5, 55, 56]. The entropy flux associated with the solar energy flux defined by Eqs. 3-4 is as follows [56, 57]:

$$S(T,\mu) = k_B \pi \sin^2 \theta_m \int_0^\infty \alpha(E)\left[(1+N_p)\ln(1+N_p) - N_p \ln N_p\right] dE. \tag{9}$$

The maximum 85% efficiency of the TPV converter can only be reached in the case of the thermal absorber-emitter connected to an ideal Carnot engine, which converts heat into free energy (i.e., electrical energy) without production of entropy [1]. However, in the process of PV energy conversion, the entropy is only conserved if an ideal PV cell is operated at (or almost at) open-circuit voltage. This condition is fulfilled in TPV systems with an ideal monochromatic filter or in the infinite-junction PV cells. Operation of a PV cell at the point of maximum efficiency is, however, an irreversible thermodynamic process accompanied by entropy creation [1, 6, 19, 56].

To estimate the entropy creation rate in the PV cell, we first calculate the energy lost in the cell via energy balance calculations:

$$dI_{PV} = I_{abs}^{E>E_g}(T_s,0) + I_u^B(T_u,0) - I_c^B(T_c,eV) - I_c^F(T_c,eV) - J \cdot V. \tag{10}$$

The difference $dI_{PV}$ between the energy carried in by the solar radiation, and the energy either emitted as a result of the radiative recombination or delivered to the load in the form of the entropy-free electrical energy is ultimately lost in the PV cell as heat. The heat dissipation is accompanied by the change in the entropy in the PV cell $dS_{PV} = dI_{PV}/T_c$. By applying the thermodynamics Second Law, we can calculate the rate of the irreversible entropy creation in the PV cell $S_{irr}^c$:

$$dS_{PV} = dI_{PV}/T_c = S_s^{E>E_g}(T_s,0) + S_u^B(T_u,0) - S_c^B(T_c,eV) - S_c^F(T_c,eV) + S_{irr}^c. \quad (11)$$

Equation 11 represents a balance between the entropy carried into the system by the absorbed solar flux, the entropy carried away in the process of radiative recombination, and the irreversible entropy created in the PV cell. On the other hand, no net energy change occurs in the up-converter: $dI_{UC} = 0$ (1), and thus, $dS_{UC} = dI_{UC}/T_u = 0$. The rate of the irreversible entropy creation in the up-converter then follows from the entropy balance:

$$dS_{UC} = 0 = S_s^{E_m<E<E_g}(T_s,0) + S_c^B(T_c,eV) - S_u^B(T_u,0) - S_u^F(T_u,0) + S_{irr}^u. \quad (12)$$

For a reversible process, both $S_{irr}^c$ and $S_{irr}^u$ would have zero values.

The rate of the entropy creation in a single-junction PV cell is shown in Fig. 8 as a function of the cell bandgap (grey dash line). The rate grows sharply for narrow gaps, reflecting large energy losses due to thermalization of electron-hole pairs, which ultimately brings down the efficiency in the narrow-gap limit. For the PV cells with progressively larger gaps, the entropy creation reduces and eventually tends to zero, however, the efficiency still declines (see e.g. Fig. 3a) due to the cell inability to utilize the sub-bandgap photons. The entropy creation rate in the PV cell as a part of an ideal hybrid converter with parameters as in Fig. 3 (blue solid lines) follows the same trend, albeit at a higher level owing to the larger energy and entropy fluxes passing through the cell. For ideal angular and spectral characteristics of the up-converter surfaces, the entropy creation rate in the up-converter is much lower than that in the PV cell, which explains high maximum efficiency levels shown in Fig. 3a. However, the increase of the radiation losses from the up-converter with non-ideal characteristics such as those in Fig. 6a,b results in the increase of the entropy creation rate in the up-converter, which overcomes the entropy creation in the cell (red solid lines in Fig. 8). This is reflected in the overall drop in the hybrid device efficiency observed in Fig. 6a.

While this work provides theoretical estimates of the maximum intrinsic efficiency limits of the solar energy conversion in a hybrid system with thermal upconversion, recent advances in nanofabrication technologies put the fabrication of selective surfaces within reach. There are on-going research efforts to accomplish angular and spectral selectivity simultaneously by the absorber surface design (Fig. 9a). Promising approaches include the use of photonic crystals [9, 13, 51, 58, 59], nano-cones, domes and pyramids [12, 13, 60], quasi-periodic planar arrays [61, 62], and plasmonic metamaterials [16, 28, 63, 64]. A simpler and cheaper solution, which does not require sophisticated design and precise nanofabrication, is to enclose the up-converter (or both, the up-converter and the PV cell) into the reflective cavity with a limited angular aperture [54] (Fig. 8b). In this case, photons emitted by the surface at larger angles will be reflected from the cavity inner surface and re-directed back to the absorber. It should be noted, however, tat as the high efficiency of the proposed conversion scheme is achieved via strong angular-selective emission properties of the up-converter surface, reduction or absence of direct sunlight will significantly degrade its performance. The actual level of efficiency reduction will depend on a specific realization of the hybrid converter and location-specific atmospheric conditions. In general, dealing with highly scattered atmospheric light due to particulates or clouds at different locations is nearly impossible to standardize, which is why we effectively ignore highly scattered light by using AM1.5 Direct (+ Circumsolar). This spectrum, which is the standard of the concentrated solar power industry, includes the direct beam from the sun plus the circumsolar component, has an integrated power density of 900 W/m$^2$, and is representative of average sunny conditions in the 48 contiguous states of the United States.

The configuration of the hybrid converter explored in detail in this paper requires two spectrally-separated beams to illuminate the converter from two opposite directions (see e.g. a schematic in Fig. 9c). Another possible configuration can be based on using two PV cells – with a thermal up-converter coupled to one of them – and one spectral splitter as shown in Fig. 9d. In specific practical realizations of the proposed conceptual conversion scheme, dichroic mirrors [65, 66] and prisms [67] can be used to achieve geometrical and spectral separation of the incoming sunlight. The former include low-pass and high-pass periodic Rugate filters, aperiodic edge filters and 3D photonic crystals such as artificial opals. Such filters have already been extensively studied and used to achieve spectral splitting in multi-junction PV systems, and can provide high optical efficiency, well-defined reflection bands, and low polarization dependence. Recently, spectrum-splitting systems based on volume transmission holographic lenses [68], coupled reflectors [66, 69], and sets of prisms [66] have also been proposed and demonstrated. They may be designed to simultaneously provide light concentration and spectral splitting. Finally, efficient absorption or reflection below the PV bandgap and efficient transmission above the bandgap can be achieved by using so-called 'plasma' filters. These filters exploit infrared plasmon excitation of doped metal oxides commonly used in the semiconductor industry such as Indium-doped Tin Oxide (ITO) or aluminum-doped Zink Oxide (AZO) [70]. By controlling the level of doping of these materials it is possible to change their plasma frequency and to accomplish efficient and spectrally-tailored low-energy absorption. To keep a focus on the estimates of the intrinsic rather than extrinsic efficiency limits of the proposed hybrid converter, we do not compare specific realizations of geometrical beam splitting.

Finally, if the up-converter can be operated at moderate temperatures, high-quality semiconductors can replace tungsten, which is conventionally used as the absorber material in TPV technology. They can provide close-to-ideal emittance characteristics of the up-converter back surface, by blocking the below-bandgap radiation. Successful realization of the developed hybrid platform offers new opportunities for reaching higher levels of the solar energy conversion efficiency under low levels of optical concentration reachable via simple low-cost technological solutions.

**Acknowledgments**

This research was supported as part of the Solid State Solar-Thermal Energy Conversion Center ($S^3$TEC), an Energy Frontier Research Center funded by the U.S. Department of Energy (DOE), Office of Science, Basic Energy Sciences (BES), under Award # DE-SC0001299/DE-FG02-09ER46577 (thermal up-conversion concept) and by the US Department of Energy under the SunShot grant # 6924527 (S-Q efficiency calculations of PV cells).

**References**

[1] P. Wurfel, Physics of solar cells: from principles to new concepts, WILEY-VCH Verlag GmbH &Co. KGaA, Weinheim, 2005.

[2] W. Shockley, H.J. Queisser, Detailed balance limit of efficiency of p-n junction solar cells, J. Appl. Phys., 32 (1961) 510.

[3] M. Green, Energy, entropy and efficiency, in: T. Kamiya, B. Monemar, H. Venghaus (Eds.) Third Generation Photovoltaics, Springer Berlin Heidelberg, 2006, pp. 21-34.

[4] A. Luque, A. Martí, Theoretical limits of photovoltaic conversion, Handbook of Photovoltaic Sci. Eng., John Wiley & Sons, Ltd, 2003, pp. 113-151.

[5] P.T. Landsberg, P. Baruch, The thermodynamics of the conversion of radiation energy for photovoltaics, J. Phys. A, 22 (1989) 1911-1926.

[6] A. De Vos, H. Pauwels, On the thermodynamic limit of photovoltaic energy conversion, Appl. Phys. A: Mater. Sci. Process., 25 (1981) 119-125.

[7] E. Yablonovitch, Statistical ray optics, J. Opt. Soc. Am., 72 (1982) 899-907.

[8] Z.F. Yu, A. Raman, S.H. Fan, Fundamental limit of nanophotonic light trapping in solar cells, Proc. Natl. Acad. Sci. USA, 107 (2010) 17491-17496.

[9] M. Florescu, H. Lee, I. Puscasu, M. Pralle, T. Lucia, Z. David, J.P. Dowling, Improving solar cell efficiency using photonic band-gap materials, Sol. Energy Mat. Sol. Cells, 91 (2007) 12.

[10] P. Spinelli, M. Hebbink, R. de Waele, L. Black, F. Lenzmann, A. Polman, Optical impedance matching using coupled plasmonic nanoparticle arrays, Nano Lett., 11 (2011) 1760-1765.

[11] H.A. Atwater, A. Polman, Plasmonics for improved photovoltaic devices, Nature Mater., 9 (2010) 205-213.

[12] S.E. Han, G. Chen, Toward the Lambertian limit of light trapping in thin nanostructured silicon solar cells, Nano Lett., 10 (2010) 4692-4696.

[13] E. Rephaeli, S. Fan, Absorber and emitter for solar thermo-photovoltaic systems to achieve efficiency exceeding the Shockley-Queisser limit, Opt. Express, 17 (2009) 15145-15159.

[14] D.M. Callahan, J.N. Munday, H.A. Atwater, Solar cell light trapping beyond the ray optic limit, Nano Lett., 12 (2012) 214-218.

[15] R.A. Pala, J. White, E. Barnard, J. Liu, M.L. Brongersma, Design of plasmonic thin-film solar cells with broadband absorption enhancements, Adv. Mater., 21 (2009) 3504-3509.

[16] Y. Yu, V.E. Ferry, A.P. Alivisatos, L. Cao, Dielectric core-shell optical antennas for strong solar absorption enhancement, Nano Lett., 12 (2012) 3674-3681.

[17] S.B. Mallick, N.P. Sergeant, M. Agrawal, J.-Y. Lee, P. Peumans, Coherent light trapping in thin-film photovoltaics, MRS Bulletin, 36 (2011) 453-460.

[18] W. Ruppel, P. Wurfel, Upper limit for the conversion of solar energy, IEEE Trans. Electron Dev., 27 (1980) 877-882.

[19] C.H. Henry, Limiting efficiencies of ideal single and multiple energy gap terrestrial solar cells, J. Appl. Phys., 51 (1980) 4494-4500.

[20] A. Marti, G.L. Araujo, Limiting efficiencies for photovoltaic energy conversion in multigap systems, Sol. Energy Mater. Sol. Cells, 43 (1996) 203-222.

[21] ASTM International West Conshohocken PA, ASTM Standard G173-03: Standard tables for reference solar spectral irradiances: direct normal and hemispherical on 37° tilted surface, 2008.

[22] A. Luque, A. Marti, Increasing the efficiency of ideal solar cells by photon induced transitions at intermediate levels, Phys. Rev. Lett., 78 (1997) 5014-5017.

[23] S.P. Bremner, M.Y. Levy, C.B. Honsberg, Limiting efficiency of an intermediate band solar cell under a terrestrial spectrum, Appl. Phys. Lett., 92 (2008) 171110-171113.

[24] P. Campbell, M.A. Green, The limiting efficiency of silicon solar cells under concentrated sunlight, IEEE Trans. Electron Devices, 33 (1986) 234-239.

[25] Z.R. Abrams, A. Niv, X. Zhang, Solar energy enhancement using down-converting particles: A rigorous approach, J. Appl. Phys., 109 (2011) 9.

[26] Z.R. Abrams, M. Gharghi, A. Niv, C. Gladden, X. Zhang, Theoretical efficiency of 3rd generation solar cells: Comparison between carrier multiplication and down-conversion, Sol. Energy Mater. Sol. Cells, 99 (2012) 308-315.

[27] T. Trupke, M.A. Green, P. Wurfel, Improving solar cell efficiencies by down-conversion of high-energy photons, J. Appl. Phys., 92 (2002) 1668-1674.

[28] C.H. Wu, B. Neuner, J. John, A. Milder, B. Zollars, S. Savoy, G. Shvets, Metamaterial-based integrated plasmonic absorber/emitter for solar thermo-photovoltaic systems, J. Opt., 14 (2012) 7.

[29] V. Badescu, Thermodynamic theory of thermophotovoltaic solar energy conversion, J. Appl. Phys., 90 (2001) 6476-6486.

[30] N.-P. Harder, P. Wurfel, Theoretical limits of thermophotovoltaic solar energy conversion, Semicond. Sci. Tech., 18 (2003) S151.

[31] N. Harder, P. , M.A. Green, Thermophotonics, Semicond. Sci. Technol., 18 (2003) S270.

[32] A. Luque, A. Marti, Limiting efficiency of coupled thermal and photovoltaic converters, Sol. Energy Mat. Sol. Cells, 58 (1999) 147-165.

[33] A.S. Vlasov, V.P. Khvostikov, O.A. Khvostikova, P.Y. Gazaryan, S.V. Sorokina, V.M. Andreev, TPV systems with solar powered tungsten emitters, AIP Conf. Proc., 890 (2007) 327-334.

[34] I. Tobias, A. Luque, Ideal efficiency and potential of solar thermophotonic converters under optically and thermally concentrated power flux, IEEE Trans. Electron Dev., 49 (2002) 2024-2030.

[35] V. Badescu, Upper bounds for solar thermophotovoltaic efficiency, Renewable Energy, 30 (2005) 211-225.

[36] T.K. Chaudhuri, A solar thermophotovoltaic converter using Pbs photovoltaic cells, Int. J. Energy Research, 16 (1992) 481-487.

[37] W. Spirkl, H. Ries, Solar thermophotovoltaics: An assessment, J. Appl. Phys., 57 (1985) 4409-4414.

[38] R.M. Swanson, A proposed thermophotovoltaic solar energy conversion system, Proc. IEEE, 67 (1979) 446-447.

[39] T.J. Coutts, M.W. Wanlass, J.S. Ward, S. Johnson, A review of recent advances in thermophotovoltaics, Photovoltaic Specialists Conf., IEEE1996, pp. 25-30.

[40] J. van der Heide, N.E. Posthuma, G. Flamand, W. Geens, J. Poortmans, Cost-efficient thermophotovoltaic cells based on germanium substrates, Solar Energy Mat. Solar Cells, 93 (2009) 1810-1816.

[41] A. Luque, V.M. Andreev, Concentrator photovoltaics, Springer, Berlin :, 2007.

[42] V. Badescu, An extended model for upconversion in solar cells, J. Appl. Phys., 104 (2008) 113120-113110.

[43] T. Trupke, M.A. Green, P. Wurfel, Improving solar cell efficiencies by up-conversion of sub-band-gap light, J. Appl. Phys., 92 (2002) 4117-4122.

[44] A.C. Atre, J.A. Dionne, Realistic upconverter-enhanced solar cells with non-ideal absorption and recombination efficiencies, J. Appl. Phys., 110 (2011) 034505-034509.

[45] S. Fischer, J.C. Goldschmidt, P. Loper, G.H. Bauer, R. Bruggemann, K. Kramer, D. Biner, M. Hermle, S.W. Glunz, Enhancement of silicon solar cell efficiency by upconversion: Optical and electrical characterization, J. Appl. Phys., 108 (2010) 044912-044911.

[46] M. Pollnau, D.R. Gamelin, S.R. Luthi, H.U. Gudel, M.P. Hehlen, Power dependence of upconversion luminescence in lanthanide and transition-metal-ion systems, Phys. Rev. B, 61 (2000) 3337-3346.

[47] T.N. Singh-Rachford, F.N. Castellano, Photon upconversion based on sensitized triplet-triplet annihilation, Coord. Chem. Rev., 254 (2010) 2560-2573.

[48] S. Baluschev, V. Yakutkin, G. Wegner, B. Minch, T. Miteva, G. Nelles, A. Yasuda, Two pathways for photon upconversion in model organic compound systems, J. Appl. Phys., 101 (2007) 023101-023104.

[49] W. Zou, C. Visser, J.A. Maduro, M.S. Pshenichnikov, J.C. Hummelen, Broadband dye-sensitized upconversion of near-infrared light, Nat Photon, 6 (2012) 560-564.

[50] A. Shalav, B.S. Richards, T. Trupke, K.W. Kramer, H.U. Gudel, Application of $NaYF_4:Er^{3+}$ up-converting phosphors for enhanced near-infrared silicon solar cell response, Appl. Phys. Lett., 86 (2005) 013505-013503.

[51] V. Badescu, Spectrally and angularly selective photothermal and photovoltaic converters under one-sun illumination, J. Phys. D: Appl. Phys., 38 (2005) 2166.

[52] M.J. Blanco, J.G. Martin, D.C. Alarcon-Padilla, Theoretical efficiencies of angular-selective non-concentrating solar thermal systems, Solar Energy, 76 (2004) 683-691.

[53] D.L.C. Chan, M. Soljacic, J.D. Joannopoulos, Thermal emission and design in 2D-periodic metallic photonic crystal slabs, Opt. Express, 14 (2006) 8785-8796.

[54] S.V. Boriskina, D. Kraemer, K. McEnaney, L.A. Weinstein, G. Chen, Solar power conversion system with directionally- and spectrally-selective properties based on a reflective cavity, U.S. Application No.: 61/697478, 2012.

[55] P.T. Landsberg, G. Tonge, Thermodynamic energy conversion efficiencies, J. Appl. Phys., 51 (1980) R1-R20.

[56] A. De Vos, Endoreversible thermodynamics of solar energy conversion, Oxford University Press, Oxford, 1992.

[57] L. Landau, On the thermodynamics of photoluminescence, J. Phys. (Moscow), 10 (1946) 503-506.

[58] M. Ghebrebrhan, P. Bermel, Y.X. Yeng, I. Celanovic, M. Soljacic, J.D. Joannopoulos, Tailoring thermal emission via Q matching of photonic crystal resonances, Phys. Rev. A, 83 (2011) 033810.

[59] P. Bermel, M. Ghebrebrhan, W. Chan, Y.X. Yeng, M. Araghchini, R. Hamam, C.H. Marton, K.F. Jensen, M. Soljacic, J.D. Joannopoulos, S.G. Johnson, I. Celanovic, Design and global optimization of high-efficiency thermophotovoltaic systems, Opt. Express, 18 (2010) A314-A334.

[60] J. Zhu, Z. Yu, S. Fan, Y. Cui, Nanostructured photon management for high performance solar cells, Mat. Sci. Eng. R: Reports, 70 (2010) 330-340.

[61] S.V. Boriskina, S.Y.K. Lee, J.J. Amsden, F.G. Omenetto, L. Dal Negro, Formation of colorimetric fingerprints on nano-patterned deterministic aperiodic surfaces, Opt. Express, 18 (2010) 14568-14576.

[62] L. Dal Negro, S.V. Boriskina, Deterministic aperiodic nanostructures for photonics and plasmonics applications, Laser Photon. Rev., 6 (2012) 178-218.

[63] C. Fu, Z. Zhang, Thermal radiative properties of metamaterials and other nanostructured materials: A review, Frontiers Energy Power Eng. China, 3 (2009) 11-26.

[64] W. Ahn, S.V. Boriskina, Y. Hong, B.M. Reinhard, Electromagnetic field enhancement and spectrum shaping through plasmonically integrated optical vortices, Nano Lett., 12 (2012) 219-227.

[65] K. Xiong, S. Lu, J. Dong, T. Zhou, D. Jiang, R. Wang, H. Yang, Light-splitting photovoltaic system utilizing two dual-junction solar cells, Solar Energy, 84 (2010), 1975-1978.

[66] A. Mojiri, R. Taylor, E. Thomsen, G. Rosengarten, Spectral beam splitting for efficient conversion of solar energy—A review, Renew. Sustain. Energy Rev. 28 (2013) 654–663.

[67] A. Barnett, D. Kirkpatrick, C. Honsberg, et al, Very high efficiency solar cell modules, Progr. Photovoltaics: Res. Appl., 17 (2008) 75-83.

[68] D. Zhang, M. Gordon, J. M. Russo, S. Vorndran, and R. K. Kostuk, Spectrum-splitting photovoltaic system using transmission holographic lenses, J. Photon. Energy, 3 (2013), 034597.

[69] D. Vincenzi, A. Busato, M. Stefancich, and G. Martinelli, Concentrating PV system based on spectral separation of solar radiation, Phys. Status Solidi A, 206, 2 (2009), 375-378.

[70] G.V. Naik, J. Kim, and A. Boltasseva, Oxides and nitrides as alternative plasmonic materials in the optical range, Opt. Mater. Express, 1, 6 (2011), 1090-1099.

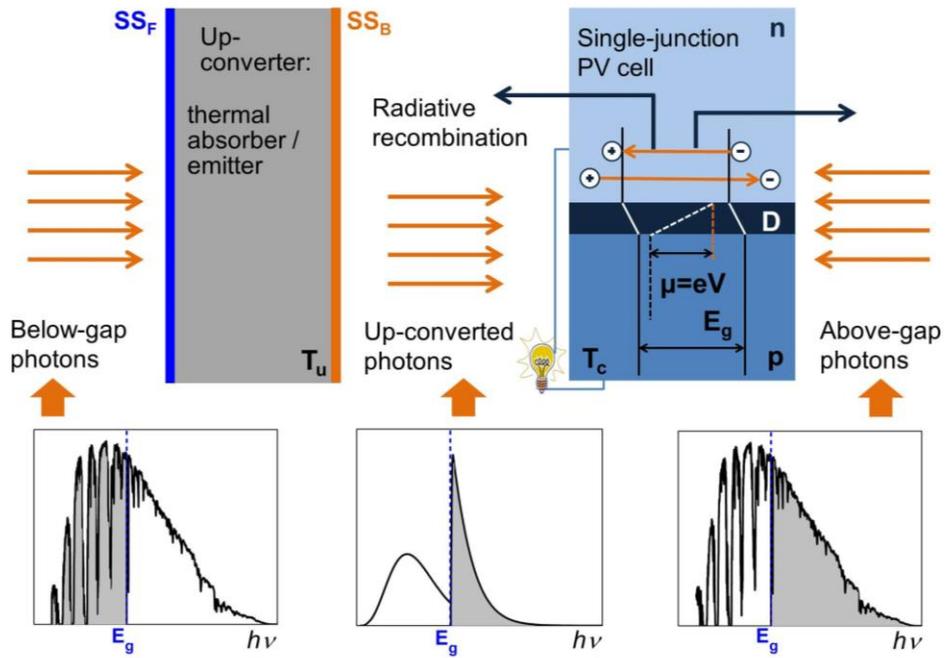

**Figure 1. Schematic of the thermal up-conversion concept.** PV cell receives sunlight photons with the energies above the bandgap through one surface, and the above-bandgap photons emitted by the up-converter through the other surface. The up-converter is a solar absorber/emitter with both front ($SS_F$) and back ($SS_B$) surfaces having spectrally- and angularly-selective emittance properties. The insets show the energy spectra of the terrestrial solar radiation (AM1.5D) and the sculpted thermal spectrum of the radiation emitted from the up-converter back surface.

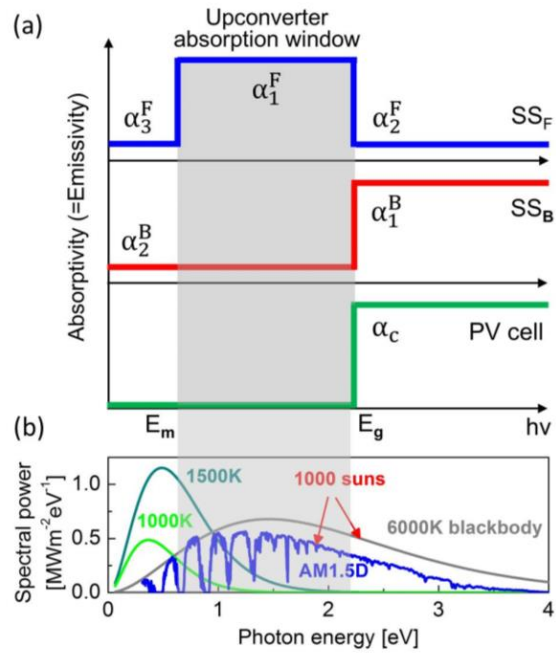

**Figure 2. Spectrally-selective surfaces**. (a) Step-wise absorption/emission characteristics of the spectrally-selective surfaces of the up-converter and the PV cell in Fig. 1. (b) The AM1.5D spectrum and the thermal radiation spectra at high temperatures. The approximation of the solar spectrum as the 6000K angularly-restricted blackbody spectrum is included for reference.

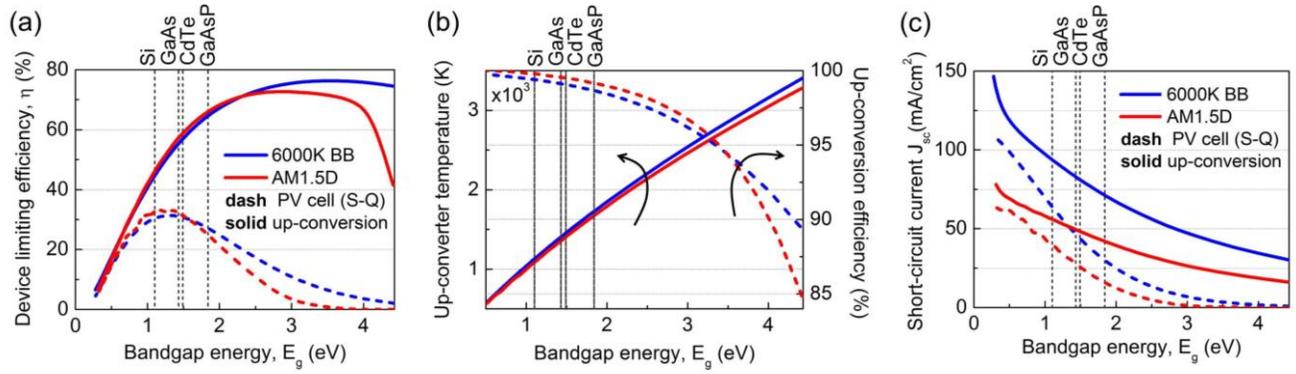

**Figure 3. Efficiency and I-V characteristics of the ideal platform**. The maximum overall efficiency and the I-V characteristics of the hybrid energy converter as a function of the PV cell bandgap energy in the ideal case: $\theta_u^F = \theta_c^F = \theta_s^m = 0.267°$, $E_m = E_s^{\min} = 0.31 eV$, $\alpha_1^F = \alpha_1^B = \alpha_c = 1$, $\alpha_2^F = \alpha_3^F = \alpha_2^B = 0$, $C=1$. (a) The efficiency of the hybrid converter for the AM1.5D spectrum (red) and a 6000K blackbody spectrum (blue). The S-Q efficiency plots for a single-junction PV cell are shown as dashed lines for comparison. The dotted lines mark the bandgap energies of several popular PV materials (labeled on top). (b) Equilibrium temperature of the up-converter (solid lines) and the up-conversion efficiency (dashed lines). (c) Short-circuit current of the PV cell (dashed lines) and the hybrid device (solid lines).

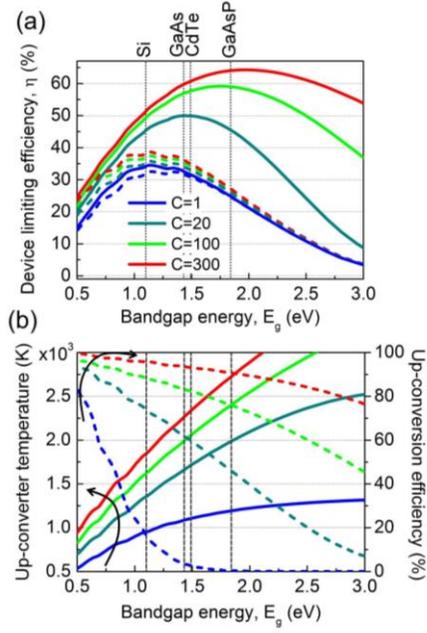

**Figure 4. Overcoming thermal losses with optical concentration**. (a) The efficiency of the hybrid energy converter as a function of the PV cell bandgap energy and the optical concentration ($\theta_u^F = 5^o$, $\theta_c^F = 90^o$, $E_m = E_s^{\min} = 0.31 eV$, $\alpha_1^F = \alpha_1^B = \alpha_c = 1$, $\alpha_2^F = \alpha_3^F = \alpha_2^B = 0$). The S-Q efficiency plots for a single-junction PV cell are shown as dashed grey lines. (b) The up-conversion efficiency (dashed lines) and equilibrium temperature of the up-converter (solid lines).

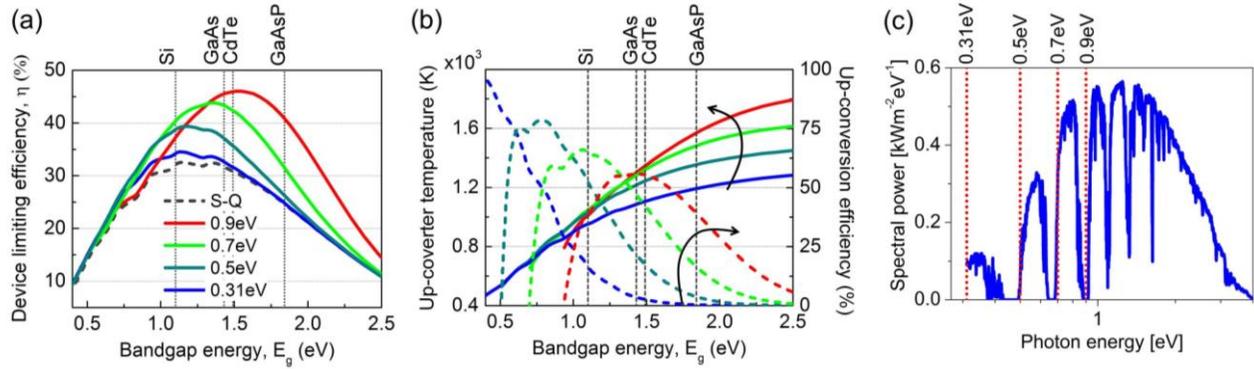

**Figure 5**. **Overcoming thermal losses with spectral positioning of the up-converter absorption window.** (a) The overall efficiency of the hybrid energy converter as a function of the PV cell bandgap energy and the spectral characteristics of the up-converter surface absorptance/emittance ($\theta_u^F = 5^o$, $\theta_c^F = 90^o$, $\alpha_1^F = \alpha_1^B = \alpha_c = 1$, $\alpha_2^F = \alpha_3^F = \alpha_2^B = 0$, C=1). The S-Q efficiency plot for a single-junction PV cell is shown as a dashed grey line. (b) The up-conversion efficiency (dashed lines) and equilibrium temperature of the up-converter (solid lines).

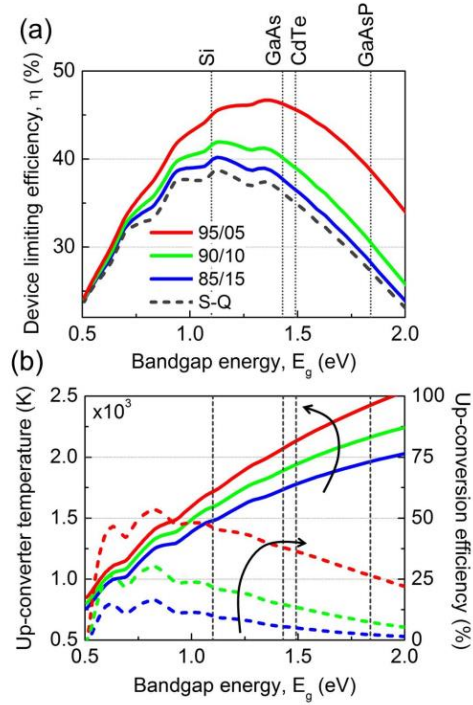

**Figure 6**. **Efficiency of the hybrid platform with non-ideal step-wise absorption characteristics.** The efficiency of the hybrid energy converter as a function of the PV cell bandgap energy and the level of radiation losses due to non-ideal selectivity of the up-converter surfaces ($\theta_u^F = 5^o$, $\theta_c^F = 90^o$, $\alpha_c = 1$, $E_m = 0.5 eV$, C=300). (a) The overall device efficiency, (b) the up-conversion efficiency (dashed lines), and the equilibrium temperature (solid lines) of the up-converter for varied absorptance/emittance characteristics of the up-converter surfaces (shown as labels). The S-Q efficiency plot for a single-junction PV cell is shown as a dashed grey line.

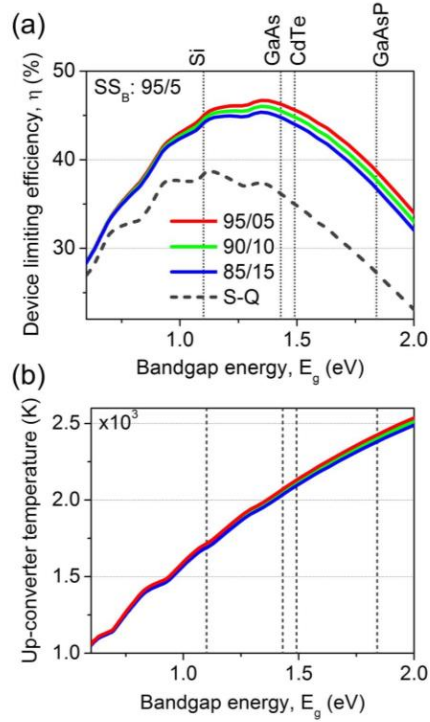

**Figure 7**. **Efficiency of the hybrid platform with non-ideal step-wise absorption characteristics.** The efficiency of the hybrid energy converter as a function of the PV cell bandgap energy and the level of radiation losses due to non-ideal selectivity of the up-converter surfaces ($\theta_u^F = 5^o$, $\theta_c^F = 90^o$, $\alpha_c = 1$, $E_m = 0.5 eV$, $C$=300). (a) The overall device efficiency, and (b) the equilibrium temperature of the up-converter for varied absorptance/emittance characteristics of the up-converter surfaces (shown as labels). The back surface losses are kept at the 95/5 level. The S-Q efficiency plot for a single-junction PV cell is shown as a dashed grey line for comparison.

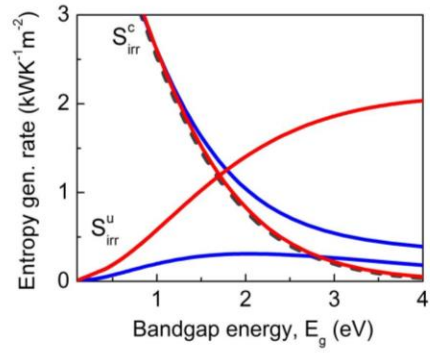

**Figure 8**. **Entropy generation in the hybrid TPV converter.** The rate of entropy creation in the PV cell $S_{irr}^c$ and in the up-converter $S_{irr}^u$ for the ideal hybrid converter with parameters as in Fig. 3 (solid blue), and for the converter with parameters as in Fig. 6a,b and 95/5 level of surface absorptance/emittance characteristics (solid red). The rate of entropy creation in a single-junction PV cell is shown as a dashed grey line for comparison.

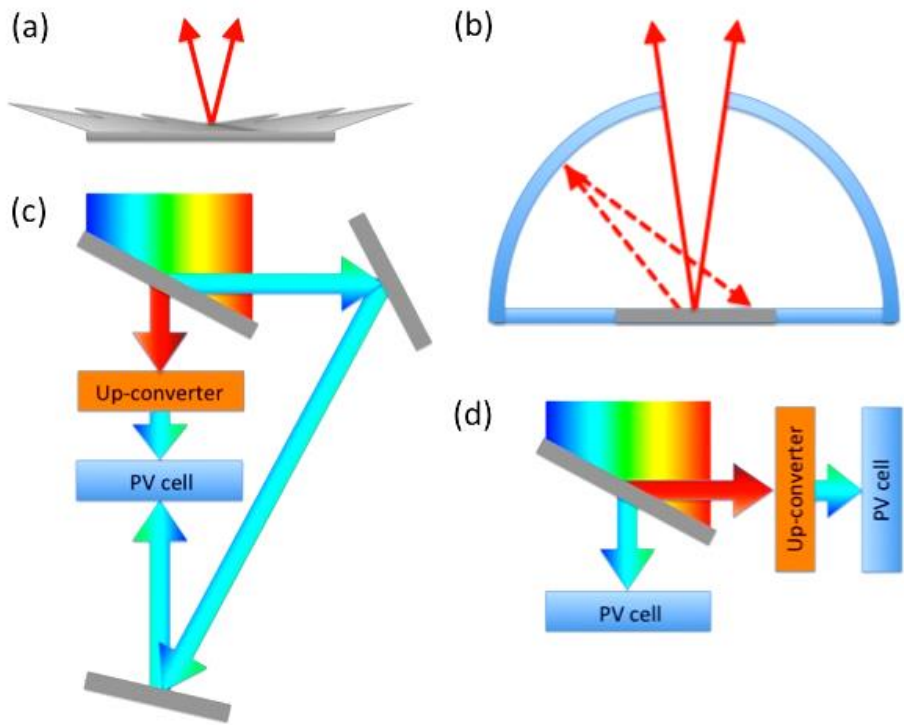

**Figure 9. Schematics of possible practical realizations of angular selectivity and spectral splitting in the proposed hybrid converter.** Angular selectivity via surface nano-patterning (a) and by enclosing the up-converter into a reflective cavity with a small aperture (b). Geometrical and spectral splitting of sunlight in a configuration with one PV cell (c) and two PV cells (d).